\documentclass{llncs}
\usepackage{makeidx}  
\usepackage{cite} 
\usepackage{hyperref}
\usepackage[utf8]{inputenc} 
\usepackage[misc]{ifsym}
\usepackage{tikz}
\usetikzlibrary{arrows,shapes,positioning,shadows,trees,calc,decorations.markings}
\begin{document}
\frontmatter          

\mainmatter              
\title{Glioma Prognosis: Segmentation of the Tumor and Survival Prediction using Shape, Geometric and Clinical Information } 
\titlerunning{BN_PixelNet}  
%
\author{Mobarakol Islam\inst{1,2} \and V Jeya Maria Jose\inst{2,3} \and
Hongliang Ren \inst{2}}
\authorrunning{Mobarakol Islam et al.} 
%
\tocauthor{Example Author} 
%
\institute{NUS Graduate School for Integrative Sciences and Engineering (NGS), National University of Singapore, Singapore
\and
Dept. of Biomedical Engineering, National University of Singapore, Singapore\\
\and
Dept. of Instrumentation and Control Engineering, National Institute of Technology, Tiruchirappalli, India\\
\email{mobarakol@u.nus.edu, jeyamariajose7@gmail.com, ren@nus.edu.sg }}

\maketitle              

\begin{abstract}
Segmentation of brain tumor from magnetic resonance imaging (MRI) is a vital process to improve diagnosis, treatment planning and to study the difference between subjects with tumor and healthy subjects. In this paper, we exploit a convolutional neural network (CNN) with hypercolumn technique to segment tumor from healthy brain tissue. Hypercolumn is the concatenation of a set of vectors which form by extracting convolutional features from multiple layers. Proposed model integrates batch normalization (BN) approach with hypercolumn. BN layers help to alleviate the internal covariate shift during stochastic gradient descent (SGD) training by zero-mean and unit variance of each mini-batch. Survival Prediction is done by first extracting features(Geometric, Fractal, and Histogram) from the segmented brain tumor data. Then, the number of days of overall survival is predicted by implementing regression on the extracted features using an artificial neural network (ANN). Our model achieves a mean dice score of  89.78\%, 82.53\% and 76.54\% for the whole tumor, tumor core and enhancing tumor respectively in segmentation task and 67.9\% in overall survival prediction task with the validation set of BraTS 2018 challenge. It obtains a mean dice accuracy of 87.315\%, 77.04\% and 70.22\% for the whole tumor, tumor core and enhancing tumor respectively in the segmentation task and a 46.8\% in overall survival prediction task in the BraTS 2018 test data set. 
\end{abstract}
\section{Introduction} 
Gliomas are the most frequent brain tumor with the highest mortality rate which develops from glial cell\cite{holland2001progenitor}. Early detection, accurate segmentation and estimation of the relative volume are very crucial for overall survival (OS) prediction, treatment and surgical planning. In addition, manual segmentation of tumor tissue is tedious, time consuming and requires strong supervision by a human expert. It is also prone to inter and intra-rater variability. So it is highly necessary to develop an automatic segmentation system to diagnose and estimate the volume, size, shape and location of the tumor. Overall survival prediction along with automatic segmentation would be a very useful tool that would help in better clinical diagnosis.

In recent years, the success of deep learning in this field is huge as it shows state of art performance in the applications of segmentation, classification, regression and detection. Khan et al. \cite{iftekharuddin2018glioblastoma} exploits convolutional neural network (CNN) for glioma segmentation and extracts handcrafted features like histogram, co-occurrence matrix, neighbourhood gray tone difference, run length, volume and areas to predict OS using random forest regression model. Alain et al.\cite{jungo2017towards} uses the residual convolutional neural network with Bayesian dropout to segment tumor and calculates the geometric features(e.g. volume, heterogeneity, rim width, surface irregularity etc.) from the segmented tumor. Later, a simple artificial neural network (ANN) is utilized to predict the exact days of OS. 3D U-net and linear regression approaches are also exploited to segment and predict OS \cite{Amorim20173D}.

In this paper, we propose a batch normalized CNN architecture  with hypercolumn features inspired by mutli-modal PixelNet \cite{islam2017multi, islam2018class, bansal2017pixelnet} where a modest number of features are extracted from multiple convolution layers and trained with a multi-layer perceptron (MLP) to predict segmentation classes. We discuss about the various features (Geometric, Fractal, Image and Clinical) extracted from the segmentation output. We propose a new method of overall survival prediction by combining all the meaningful features that contribute to the number of days of survival left for the patient. Details about how the features are selected and various experiments that are run for finding the best regression technique has  been discussed. The problems of generalizing a network over its performance on a validation data set is also discussed.

\section{Methods} 
In this section, we discuss about the dataset provided, the model and methodologies to process data for segmentation and about the extraction of features, feature selection and regression techniques that were exploited for overall survival prediction.

\subsection{Dataset}

BraTS 2018 (Brain Tumor Image Segmentation Benchmark) training database\cite{bakas2017segmentation,bakas2017advancing,menze2015multimodal,bakas2017segmentationlgg} consists in total 285 cases of patients out of which the overall survival prediction data was provided for 163 cases. BraTS 2018 Validation dataset \cite{bakas2017segmentation,bakas2017advancing,menze2015multimodal,bakas2017segmentationlgg} consists of 53 cases. All the data are a multi-modal MRI scan of 210 high-grade glioma (HGG) and 75 low-grade glioma (LGG) and 4 different modalities including T1 (spin-lattice relaxation), T1c (T1-contrasted), T2 (spin-spin relaxation) and FLAIR (fluid attenuation inversion recovery). Each scan is a continuous 3D volume of 155 2D slices of size 240x240. The volume of the various modalities is already skull-stripped, aligned with T1c and interpolated to 1 mm voxel resolution. 

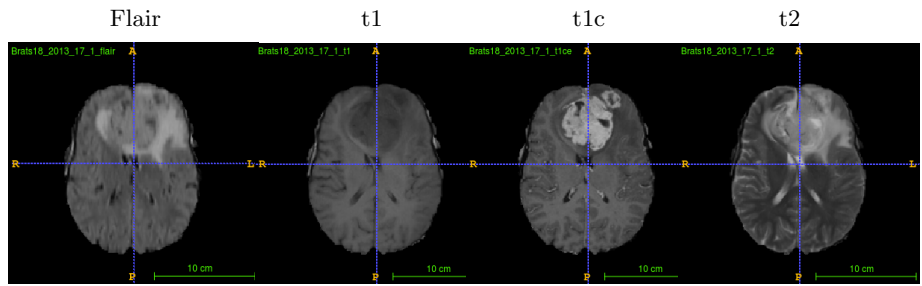
\begin{figure}[!htbp]
\begin{minipage}{.94\textwidth}
  \centering
  \begin{tikzpicture}[remember picture]
   \node [anchor=south east, xshift=-120, yshift=-0] (csupp){\includegraphics[width=.29\textwidth]{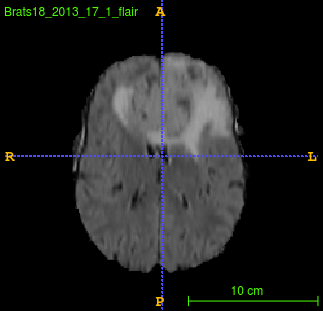}};
   \node [anchor=south east, xshift=-30, yshift=-0] (csupp){\includegraphics[width=.28\textwidth]{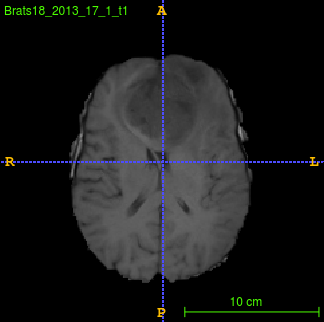}};
   \node [anchor=south east, xshift=50, yshift=-0] (csupp){\includegraphics[width=.28\textwidth]{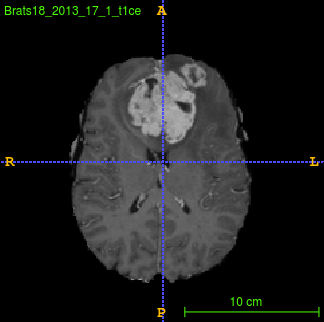}};
    \node [anchor=south east, xshift=130, yshift=-0] (csupp){\includegraphics[width=.28\textwidth]{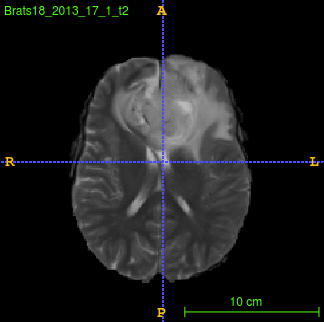}};
    
      \node[anchor=south west,xshift=-215, yshift=0] at (csupp.north west) {Flair};
   \node[anchor=south west,xshift=-120, yshift=0] at (csupp.north west) {t1};
   \node[anchor=south west,xshift=-40, yshift=0] at (csupp.north west) {t1c};
   \node[anchor=south west,xshift=38, yshift=0] at (csupp.north west) {t2};

  \end{tikzpicture}
\end{minipage}
\caption{Visualization of the different modalities in the BraTS 2018 Training data set.
\label{fig:tra}}
\end{figure}

\begin{figure}[!htbp]
\begin{minipage}{.94\textwidth}
  \centering
  \begin{tikzpicture}[remember picture]
   \node [anchor=south east, xshift=0, yshift=-100] (csupp){\includegraphics[width=.28\textwidth]{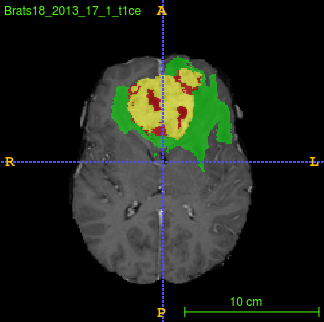}};
   \node[anchor=south west,xshift=38, yshift=0] at (csupp.north west) {GT};
  \end{tikzpicture}
\end{minipage}
\caption{Ground Truth with different segmentation labels as given in BraTS 2018 Training data set. Red, Yellow and Green colors represent Necrotic, Enhancing and Edema respectively. 
\label{fig:trainseg}}
\end{figure}

The provided ground truth with manual segmentation includes three labels: GD-enhancing tumor (ET — label 4), the peritumoral edema (ED — label 2), and the necrotic and non-enhancing tumor (NCR/NET — label 1). The predicted labels are evaluated by merging three regions: whole tumor (WT: all four labels), tumor core (TC: 1,2) and enhancing tumor (ET: 4). Fig. \ref{fig:tra} illustrates all the four modalities for one of the samples of the training data set of BraTS 2018. Fig. \ref{fig:trainseg} illustrates the provided ground truth for the same sample. The green label corresponds to GD-enhancing tumor, yellow label corresponds to peritumoral edema and the red label corresponds to necrotic and non-enhancing tumor. ITK-SNAP\cite{yushkevich2016itk} is the tool that was used to visualize the data.

\subsection{Segmentation}
\subsubsection{Proposed Model}
\begin{figure}[!htbp]
\begin{minipage}{.94\textwidth}
  \centering
  \begin{tikzpicture}[remember picture]
   \node [anchor=south east, xshift=0, yshift=0] (csupp){\includegraphics[]{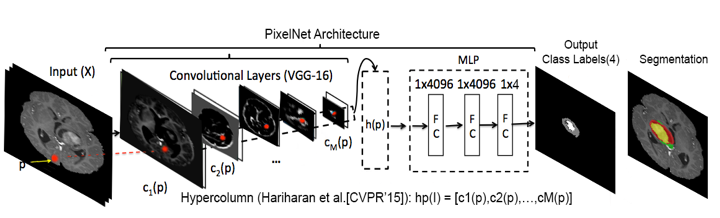}};
  \end{tikzpicture}
\end{minipage}
\caption{Batch Normalized PixelNet Architecture.
\label{fig:pixelnet}}
\end{figure}

Our proposed model is inspired from PixelNet \cite{bansal2017pixelnet, islam2017multi} where we integrate additional 3 convolution layers and batch normalization after all the convolution layers. The model consists of 18 pixel-block and a hypercolumn layer followed by a multi-layer perceptron (MLP) of 3 fully connected layers as in Fig.\ref{fig:pixelnet}. A pixel-block contains convolution, batch normalization (BN)\cite{ioffe2015batch} and ReLu layers sequentially. Hypercolumn layer extracts the features from multiple convolution layers and concatenates them into a feature vector which propagates to the MLP for pixel-wise classification.

\subsection{Survival Prediction}
Following the extraction of tumor from the MRI scans, the segmented tumor along with certain other parameters are used for survival prediction. The following paragraphs elucidate the features those are extracted along with the regression model that is built for predicting survival.

\subsubsection{Feature Extraction}
\begin{figure}[!tbp]
\begin{minipage}{.94\textwidth}
  \centering
  \begin{tikzpicture}[remember picture]
   \node [anchor=south east, xshift=-100, yshift=-0] (csupp){\includegraphics[width=.309\textwidth]{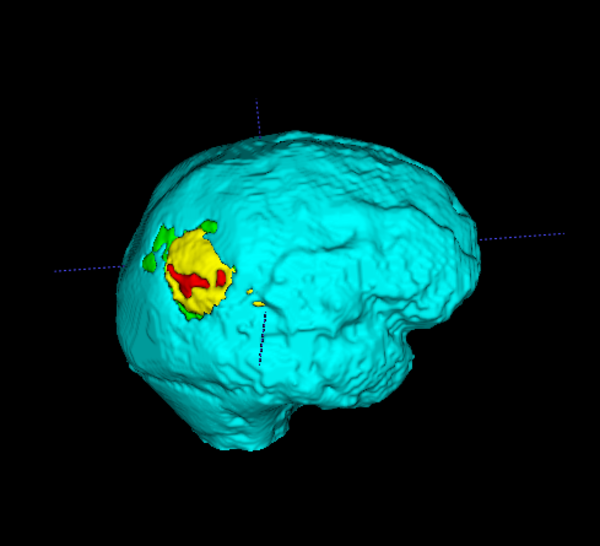}};
   \node [anchor=south east, xshift=-0, yshift=-0] (csupp){\includegraphics[width=.3\textwidth]{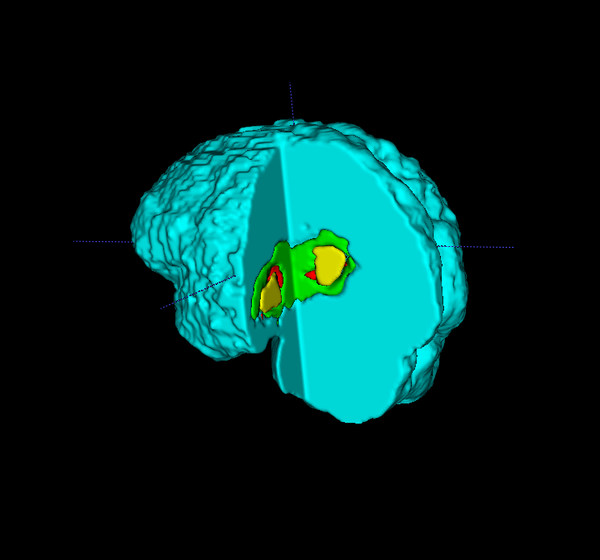}};
   \node [anchor=south east, xshift=100, yshift=-0] (csupp){\includegraphics[width=.3\textwidth]{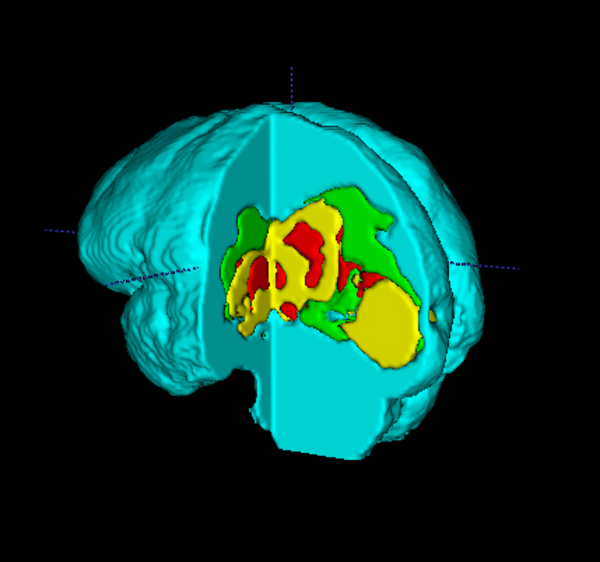}};
   \node[anchor=south west,xshift=20, yshift=-120] at (csupp.north west) {Short survival};
   \node[anchor=south west,xshift=-90, yshift=-120] at (csupp.north west) {Medium survival};
   \node[anchor=south west,xshift=-180, yshift=-120] at (csupp.north west) {Long survival};

  \end{tikzpicture}
\end{minipage}
\caption{Visualization of tumor with different survival rate.
\label{fig:survival_example}}
\end{figure}
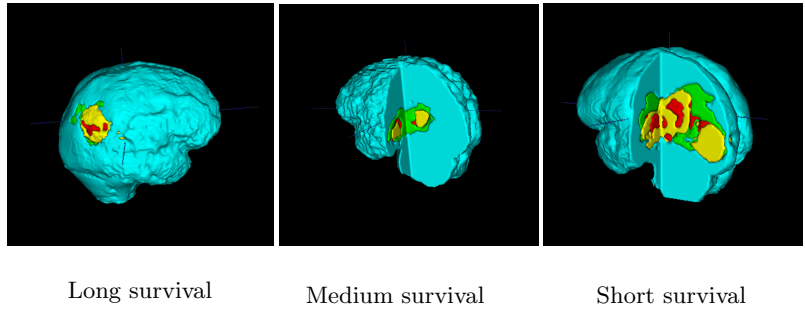

\begin{figure}[!tbp]
\begin{minipage}{.94\textwidth}
  \centering
  \begin{tikzpicture}[remember picture]
   \node [anchor=south east, xshift=-100, yshift=-0] (csupp){\includegraphics[width=.5\textwidth]{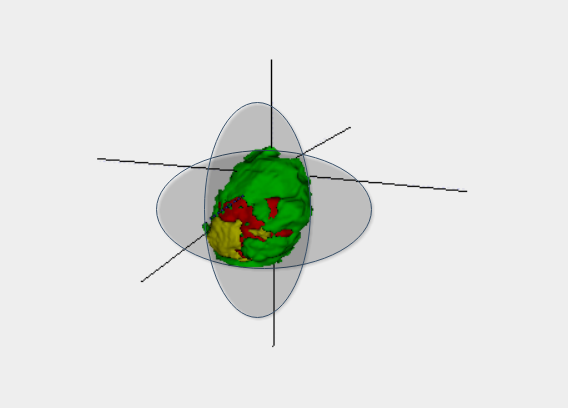}};

  \end{tikzpicture}
\end{minipage}
\caption{Visualization of the extracted tumor from brain.
\label{fig:geo}}
\end{figure}

The tumor geometry and its location hold a very important role in deciding the number of days of survival \cite{perez2017glioblastoma}. Fig. \ref{fig:survival_example} visually illustrates how the features such as location or centroid of tumor, size and shape of tumor affect the overall survival of the patient. It is evident that more the proximity of tumor to the centre of brain, the lower is the overall survival of patient. Also, lesser the size or smaller the shape of tumor, higher is the overall survival of patient. So, we extract geometrical features which include First axis coordinates, Second axis coordinates, Third axis coordinates,Eigen Values, First axis length, Second axis length, Third axis Length, Centroid coordinates, meridional eccentricity and equatorial eccentricity for individual tumor types as well as whole tumor. Fig. \ref{fig:geo} illustrates the tumor extracted from the brain and gives an intuition of how the geometric features like centroid, eccentricity and axis lengths are calculated. Also, the volume of tumor and its ratio with respect to the total volume was calculated. Features of the tumor image including mean, variance, standard deviation, entropy, skewness\cite{mardia1970measures}, Kurtosis\cite{mardia1970measures}, entropy and histogram feature intensities were extracted. Fractal dimension of the necrotic tumor has been found to play a pivotal role in the survival prediction, according to \cite{reishofer2018age}. So, fractal dimension and fractal ratio were extracted for the necrotic part of the segmented tumor core. In addition , the age of the subject provided by the BraTS 2018 training data set was also included.

\subsubsection{Feature Selection}
Several experiments were conducted using different combinations of extracted features. After analyzing the cross-validation errors of the experiments, the most informative features are alone retained and others are neglected. Features like eigen values, eccentricity,  skewness, mean and variance were not found to have an important role in survival prediction. Also, geometric features of the GD-enhancing tumor (ET — label 4) was found to be only increasing the cross-validation errors and hence was removed. So finally, a total of 50 features that were found to be the most informative are used in the regression model. These features are First axis coordinates, Second axis coordinates, Third axis coordinates, First  axis  length, Second  axis length, Third  axis Length, Centroid coordinates for part wise non-enhancing tumor core (NCR/NET — label 1), peritumoral edema (ED — label 2) as well as for the whole tumor without including GD-enhancing tumor (ET — label 4) in addition to Kurtosis, Entropy, Histogramic intensity, Fractal dimension and age.

\subsubsection{Regression Model}

\begin{figure}[!htbp]
\begin{minipage}{.94\textwidth}
  \centering
  \begin{tikzpicture}[remember picture]
   \node [anchor=south east, xshift=-100, yshift=-0] (csupp){\includegraphics[width=.5\textwidth]{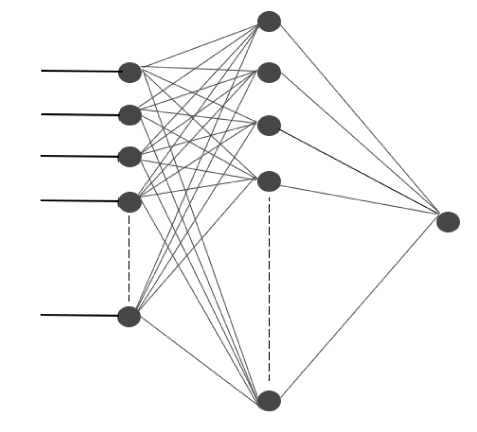}};
     \node[anchor=south west,xshift=-40, yshift=-33] at (csupp.north west) {Feature 1};
          \node[anchor=south west,xshift=-40, yshift=-48] at (csupp.north west) {Feature 2};
         \node[anchor=south west,xshift=-40, yshift=-63] at (csupp.north west) {Feature 3};    
              \node[anchor=south west,xshift=-40, yshift=-78] at (csupp.north west) {Feature 4
              };
         \node[anchor=south west,xshift=-40, yshift=-117] at (csupp.north west) {Feature 50};
         \node[anchor=south west,xshift=63, yshift=-5] at (csupp.north west) {Hidden Layer};
    \node[anchor=south west,xshift=30, yshift=-5] at (csupp.north west) {Input};
    \node[anchor=south west,xshift=133, yshift=-5] at (csupp.north west) {Output};
\node[anchor=south west,xshift=163, yshift=-87] at (csupp.north west) {Days of Survival};

  \end{tikzpicture}
\end{minipage}
\caption{Regression model using ANN.
\label{fig:ann}}
\end{figure}
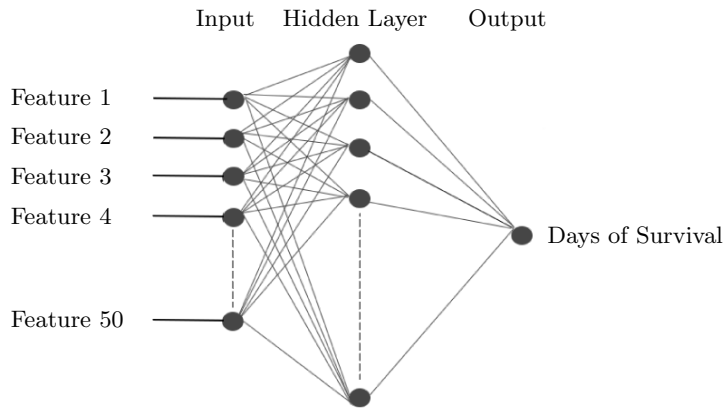

With the selected features, we train a fully connected artificial neural network (ANN) with one hidden layer\cite{schalkoff1997artificial} and ReLu activation function. Fig. \ref{fig:ann} shows the ANN model that takes 50 features as input and gives the number of days of survival as the output. We run an experiment to find which configuration of the hidden layers gives the lowest mean squared error (MSE). The best configuration was found to be 100 neurons in the hidden layer. Then, we find the best epoch by finding the epoch for which the MSE is minimum by using a cross-validation data set while training. However, after analyzing many experiments, we find that when MSE is minimum, the accuracy is low and vice-versa. So, we find an epoch where the accuracy and the MSE of the model are balanced. Adam optimizer\cite{kingma2014adam} with MSE loss function has been used to conduct all experiments with ANN. We also tune the hyper parameters like learning rate and batch size. The best result for BraTS 2018 Validation Data was acquired for the 900\textsuperscript{th} epoch with a batch size of 10.

\section{Experiments and Results}

\subsection{Segmentation}
We convert the 3D voxel of 240x240x155 into 2D slices of 240x240 by ignoring blank slices of scan and ground-truth. We choose a sample of 2000 pixels per image and batch size of 10 in training time. We normalize data to zero mean and standard daviation and augment by flipping left-right. In testing phase, hypercolum has been formed with all the pixels inside brain region and predict slice by slice to form MRI. Finally, we adopt largest component analysis to remove false positive as a post-processing technique. We utilize Caffe \cite{jia2014caffe} framework with a single Nvidia GPU 1080Ti GPU to perform all the experiments.
Table \ref{table:valid} represents the Dice and Hausdorff performance of our model. It obtains dice accuracy of 89.78\%, 82.53\% and 76.54\% of whole tumor (WT), tumor core (TC) and enhance tumor (ET) respectively. Table \ref{tabletest} shows the performance metrics that the segmentation method has achieved on the test data set. It obtains dice accuracy of 87.315\%, 77.04\% and 70.22\% of whole tumor (WT), tumor core (TC) and enhance tumor (ET) respectively.
\begin{table}[!htbp]
 \caption{Dice and Hausdorff for BraTS 2018 validation dataset}
\begin{center}
\label{table:valid}
\begin{tabular}{c|c|c|c|c|c|c}
\cline{1-7}
 & \multicolumn{3}{|c|}{Dice} & \multicolumn{3}{|c}{Hausdorff}\\
\cline{1-7}
{}      &WT     &TC     &ET     &WT     &TC     &ET \\ \hline
Mean	&\textbf{89.78}	&82.53	&76.54	&5.09   &7.11   &3.60	\\ \hline
StdDev	&8.35	&17.80	&23.23	&7.04	&8.04   &5.58\\ \hline
Median	&90.51	&86.87	&83.15	&3.08	&4.53   &2.23 \\ \hline
\end{tabular}
\end{center}
\end{table}

\begin{table}[!htbp]
 \caption{Dice and Hausdorff for BraTS 2018 Test dataset \cite{bakas2018identifying}}
\begin{center}
\label{tabletest}
\begin{tabular}{c|c|c|c|c|c|c}
\cline{1-7}
 & \multicolumn{3}{|c|}{Dice} & \multicolumn{3}{|c}{Hausdorff}\\
\cline{1-7}
{}      &WT     &TC     &ET     &WT     &TC     &ET \\ \hline
Mean	&\textbf{87.315}	&77.044	&70.22	&7.15   &8.0024   &4.594	\\ \hline
StdDev	&11.33	&26.082	&27.96	&13.08	&11.937   &8.802\\ \hline
Median	&90.36	&87.89	&80.93	&3.6	&4.24   &2.23 \\ \hline
\end{tabular}
\end{center}
\end{table}

The following Fig. \ref{fig:seg} shows the visualizations of output from the segmentation network along with the input for BraTS 2018 \cite{bakas2018identifying} Test data set. The different modalities of the input namely- Flair,t1c and t2 are also illustrated. The prediction shows the labelling of tumor done by the network. It consists of 3 labels, green corresponds to GD-enhancing tumor, yellow corresponds to peritumoral edema and red corresponds to necrotic and non-enhancing tumor core. 
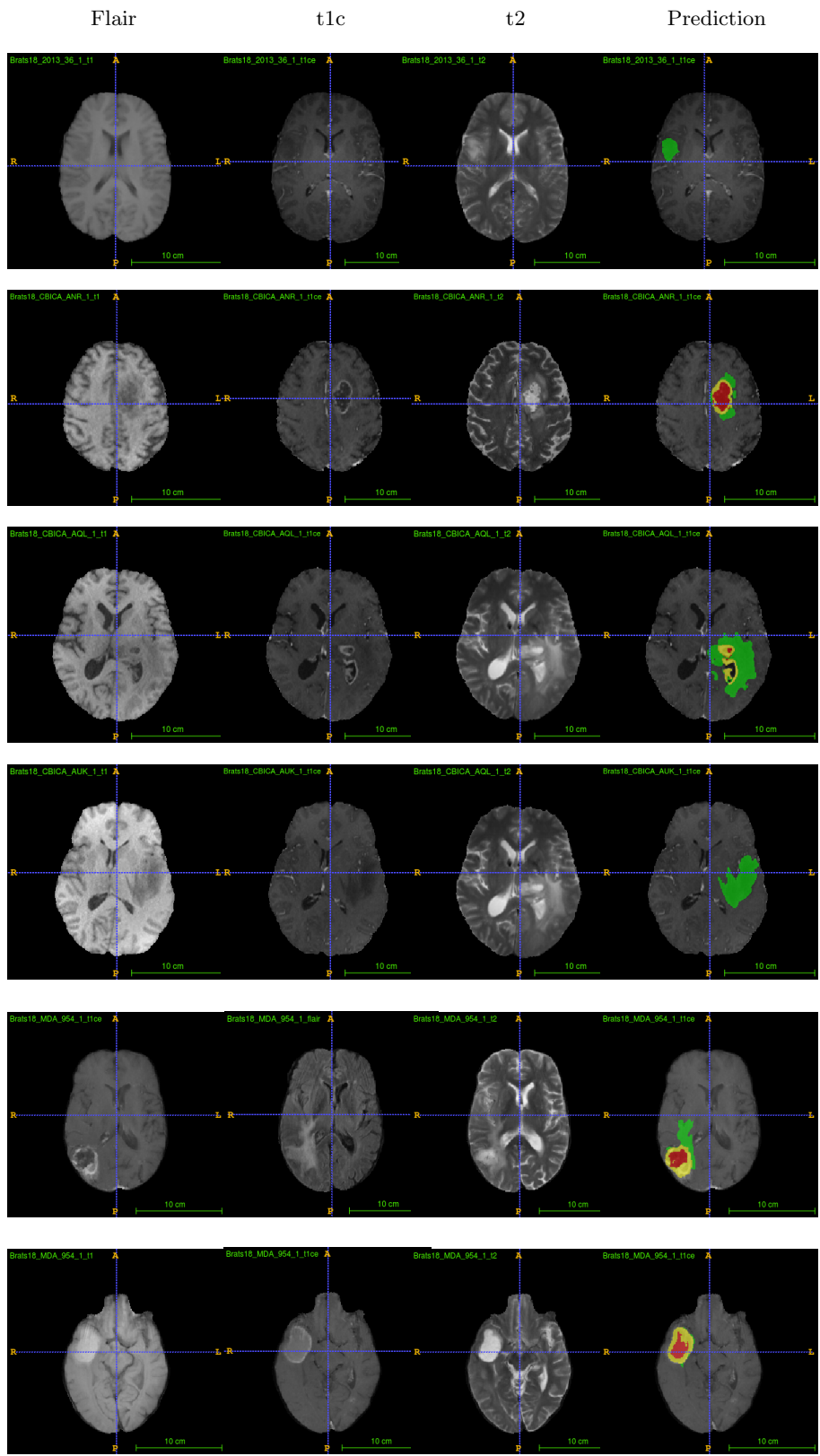
\begin{figure}[!tbp]
\begin{minipage}{.94\textwidth}
  \centering
  \begin{tikzpicture}[remember picture]
   \node [anchor=south east, xshift=-120, yshift=-0] (csupp){\includegraphics[width=.28\textwidth]{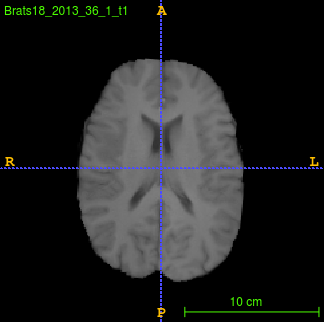}};
   \node [anchor=south east, xshift=-30, yshift=-0] (csupp){\includegraphics[width=.28\textwidth]{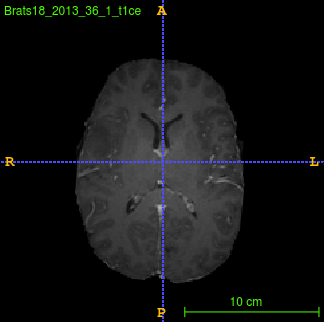}};
   \node [anchor=south east, xshift=50, yshift=-0] (csupp){\includegraphics[width=.295\textwidth]{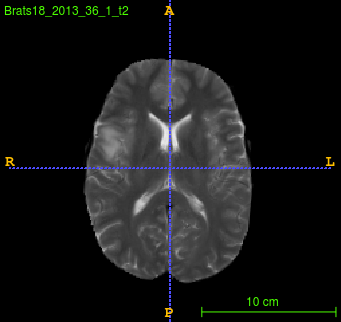}};
    \node [anchor=south east, xshift=130, yshift=-0] (csupp){\includegraphics[width=.28\textwidth]{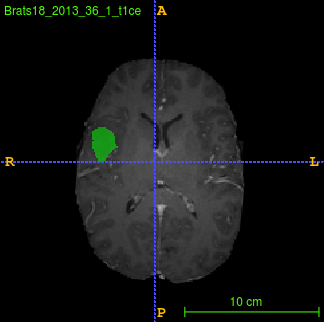}};
    
     \node [anchor=south east, xshift=-120, yshift=-100] (csupp){\includegraphics[width=.28\textwidth]{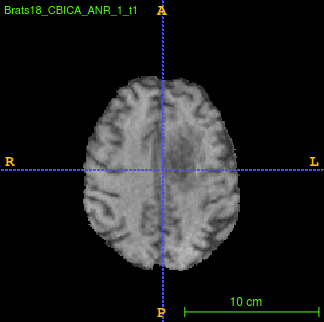}};
   \node [anchor=south east, xshift=-30, yshift=-100] (csupp){\includegraphics[width=.28\textwidth]{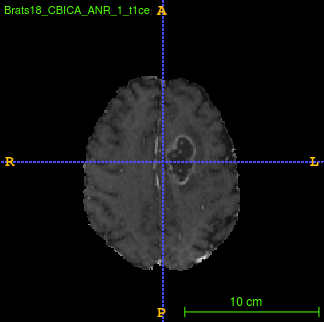}};
   \node [anchor=south east, xshift=50, yshift=-100] (csupp){\includegraphics[width=.28\textwidth]{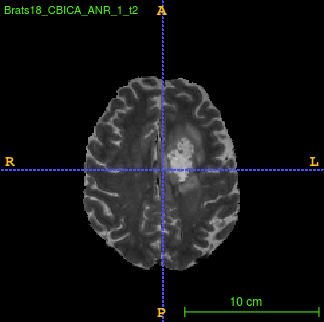}};
    \node [anchor=south east, xshift=130, yshift=-100] (csupp){\includegraphics[width=.28\textwidth]{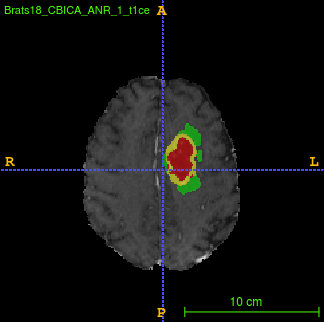}};
    
     \node [anchor=south east, xshift=-120, yshift=-200] (csupp){\includegraphics[width=.28\textwidth]{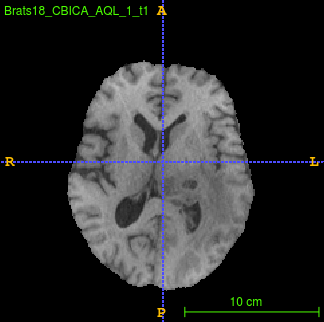}};
   \node [anchor=south east, xshift=-30, yshift=-200] (csupp){\includegraphics[width=.28\textwidth]{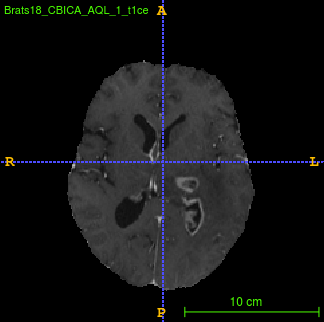}};
   \node [anchor=south east, xshift=50, yshift=-200] (csupp){\includegraphics[width=.28\textwidth]{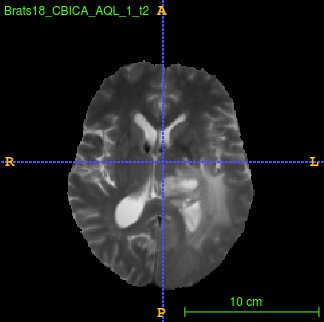}};
    \node [anchor=south east, xshift=130, yshift=-200] (csupp){\includegraphics[width=.28\textwidth]{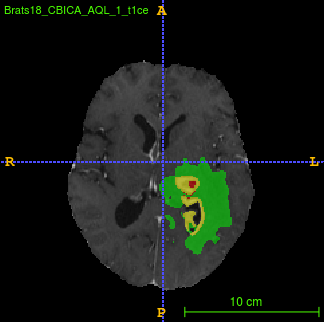}};
    
     \node [anchor=south east, xshift=-120, yshift=-300] (csupp){\includegraphics[width=.28\textwidth]{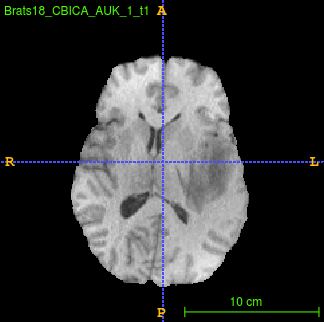}};
   \node [anchor=south east, xshift=-30, yshift=-300] (csupp){\includegraphics[width=.28\textwidth]{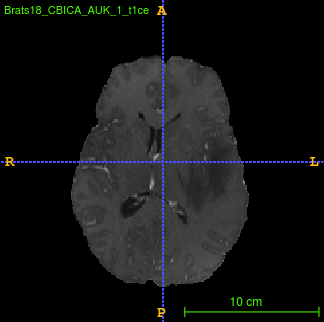}};
   \node [anchor=south east, xshift=50, yshift=-300] (csupp){\includegraphics[width=.28\textwidth]{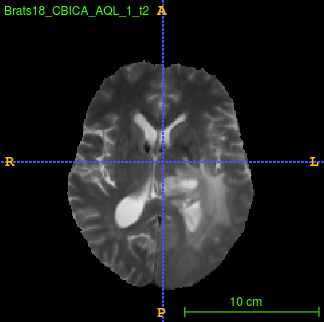}};
    \node [anchor=south east, xshift=130, yshift=-300] (csupp){\includegraphics[width=.28\textwidth]{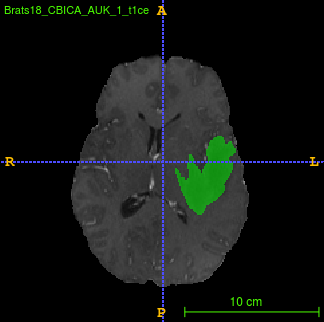}};
    
     \node [anchor=south east, xshift=-120, yshift=-400] (csupp){\includegraphics[width=.28\textwidth]{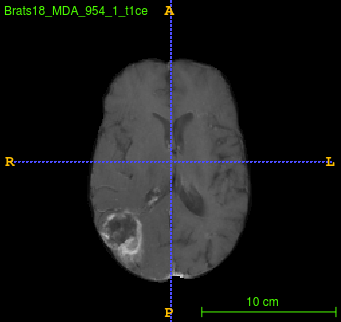}};
   \node [anchor=south east, xshift=-30, yshift=-400] (csupp){\includegraphics[width=.276\textwidth]{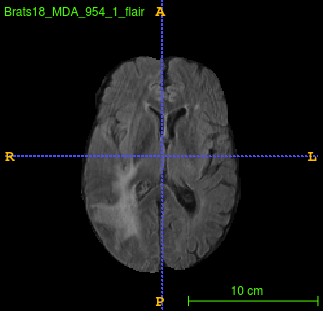}};
   \node [anchor=south east, xshift=50, yshift=-400] (csupp){\includegraphics[width=.28\textwidth]{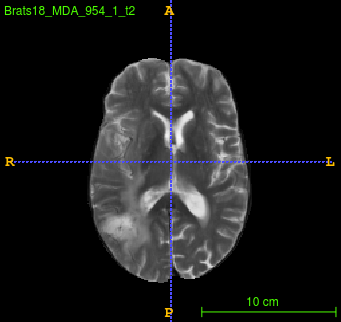}};
    \node [anchor=south east, xshift=130, yshift=-400] (csupp){\includegraphics[width=.28\textwidth]{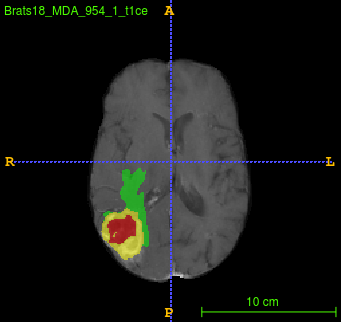}};
    
    \node [anchor=south east, xshift=-120, yshift=-500] (csupp){\includegraphics[width=.28\textwidth]{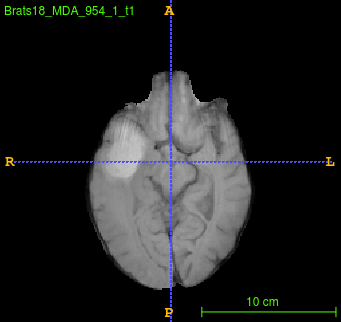}};
   \node [anchor=south east, xshift=-33, yshift=-500] (csupp){\includegraphics[width=.268\textwidth]{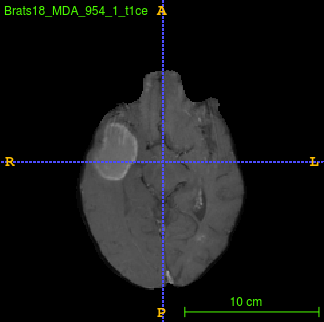}};
   \node [anchor=south east, xshift=50, yshift=-500] (csupp){\includegraphics[width=.28\textwidth]{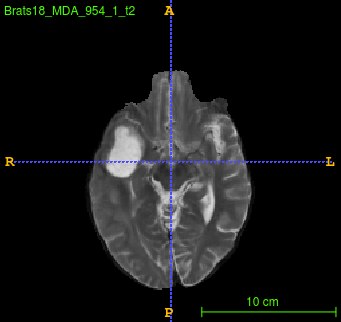}};
    \node [anchor=south east, xshift=130, yshift=-500] (csupp){\includegraphics[width=.28\textwidth]{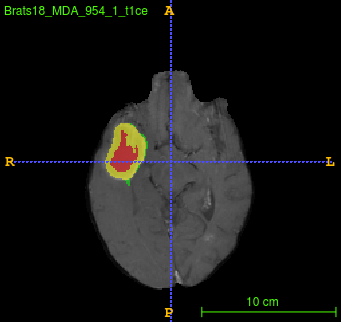}};
    
   \node[anchor=south west,xshift=-215, yshift=510] at (csupp.north west) {Flair};
   \node[anchor=south west,xshift=-120, yshift=510] at (csupp.north west) {t1c};
   \node[anchor=south west,xshift=-40, yshift=510] at (csupp.north west) {t2};
   \node[anchor=south west,xshift=28, yshift=510] at (csupp.north west) {Prediction};

  \end{tikzpicture}
\end{minipage}
\caption{Visualization of the predicted segmentation labels for BraTS 2018 Test data set.
\label{fig:seg}}
\end{figure}

\subsection{Survival Prediction}
Evaluation of the survival prediction model is performed on BraTS 2018 survival validation dataset (which is a subset (28 out of 53 subjects) of the segmentation validation dataset) \cite{bakas2017segmentation,bakas2017advancing,menze2015multimodal,bakas2017segmentationlgg}. Quantitative details for the ANN that gave the best accuracy of 67.9\%  has been listed in Table \ref{table:result_survival_rate}. This is the best accuracy that was achieved in the Validation Leader board till the time this paper is written. However, the same ANN was found to give relatively less accuracy for the test data set. Table \ref{table:result_survival_rate2} shows the quantitative results for BraTS 2018 \cite{bakas2018identifying} Test data set.

\begin{table}[!htbp]
 \caption{Quantitative results of survival prediction on the BraTS 2018 validation data set using ANN}
  \centering
\label{table:result_survival_rate}
\begin{tabular}{c|c|c|c|c|c}
\hline
Cases 	&Accuracy	&MSE	&MedianSE	&stdSE	&SpearmanR	\\ \hline
28	&\textbf{67.9\%}	&{96161.713}	&{59473.481}	&117207.189	&0.496	\\ \hline
\end{tabular}
\end{table}

\begin{table}[!htbp]
 \caption{Quantitative results of survival prediction on the BraTS 2018 Test data set using ANN}
  \centering
\label{table:result_survival_rate2}
\begin{tabular}{c|c|c|c|c|c}
\hline
Cases 	&Accuracy	&MSE	&MedianSE	&stdSE	&SpearmanR	\\ \hline
77	&\textbf{46.8\%}	&{341387.439}	&{92892.2261}	&788491.731	&0.148	\\ \hline
\end{tabular}
\end{table}

\section{Discussion}
\subsection{Experiments for finding the best regression technique}
For finding out the best regression model and the meaningful features, several experiments are conducted. Experiments with different combinations of available features and regression models are conducted on the 33 subjects of BraTS17 validation dataset \cite{bakas2017segmentation,bakas2017advancing,menze2015multimodal} to find the best possible combination of features and regression model. The details of the experiments on the entire features and the best 50 features are listed in Table \ref{fig:survival_example} and \ref{table:crossvalid} respectively. When compared to ANN, other regression models such as Support Vector Machine(SVM) with Radial basis function(RBF) kernel\cite{joachims1998making}, Random Forest\cite{liaw2002classification}, Linear Regression\cite{neter1989applied}, Logistic Regression\cite{hosmer2013applied} are investigated but resulted in inferior performance. Also, using only the best 50 features showed improvement in accuracy and MSE for every models.

\begin{table}[!htbp]
 \caption{Performance comparison with all the features in different machine learning models for BraTS 2017 validation data set.}
  \centering
\label{table:crossvalid}
\begin{tabular}{c|c|c|c|c|c}
\hline
Models 	&Accuracy	&MSE	&MedianSE	&stdSE	&SpearmanR	\\ \hline
Linear Regression 	&50.5\%	&252353.061	&95419.102	&\textbf{429879.191}	&0.263	\\ \hline
SVM	&33.3\%	&242147.277	&62044.450	&563941.194	&0.142	\\ \hline
Random Forest	&42.4\%	&\textbf{208660.63}	&\textbf{33367.111}	&502312.762	&0.213	\\ \hline
Logistic Regression	&39.4\%	&286470	&40401	&540201.363	&\textbf{0.479}\\ \hline
\textbf{ANN}	&\textbf{54.5\%}	&211967.681	&53967.305	&540221.112	&0.206	\\ \hline

\end{tabular}
\end{table}

\begin{table}[!htbp]
 \caption{Performance comparison with the best 50 features in different machine learning models for BraTS 2017 validation data set.}
 \centering
\label{table:crossvalid1}
\begin{tabular}{c|c|c|c|c|c}
\hline
Models 	&Accuracy	&MSE	&MedianSE	&stdSE	&SpearmanR	\\ \hline
Linear Regression 	&50.5\%	&237148.501	&78915.034	&362705.604	&0.221	\\ \hline
SVM	&42.4\%	&233367.604	&127524.752	&374037.279	&0.218	\\ \hline
Random Forest	&39.4\%	&262224.703	&\textbf{42507.088}	&506754.007	&\textbf{0.324}	\\ \hline
Logistic Regression	&36.4\%	&\textbf{181509.182}	&58564	&\textbf{249213.006}	&0.13\\ \hline
\textbf{ANN}	&\textbf{60.6\%}	&214207.487	&60832.523	&354332.371	&0.293	\\ \hline

\end{tabular}
\end{table}
\subsection{Issues with generalizing a network on validation data set}
The quantitative results of survival prediction on the BraTS 2018 test data set can be found in Table \ref{table:result_survival_rate2}. When we use the same algorithm with the same 50 features that gave the best results in the validation data set for the test data set of survival prediction task, we find that the accuracy of the network has reduced drastically. Also, the mean squared error obtained increases. So, it is evident that the network does not perform well with the test data set as good it performs for the validation data set. The reason for this behaviour is that the model is not generalized. It is over-fitted to the train data. 

 The same architecture, with the same number of neurons in the hidden layers, with the same hyper parameters; trained for a certain number of epochs that gave the best accuracy for validation data set was used for the test data set. The number of epochs was fixed by experimenting which epoch gave the least mean squared error for the validation data set. This is where the network was over-fitted. Generalizing a network architecture and the number of epochs till which it should be trained should not be done by comparing the results with a few data. The data with which a network is generalized can be of a very meagre amount and might contain similar characteristics. So generalizing a network over the validation data set here has caused the network to perform badly in a fresh test data set.
 
 Looking at Table \ref{table:crossvalid1}, models like Logistic regression or Random Forest which would not have over fitted the data could have been used. Thus, selection of validation data is very crucial for any neural network as well as deep learning model where the data  set is really small.   
\section{Conclusion}
Batch Normalized pixelnet is found to give quality segmentation results for the BraTS 2018 Validation data set. The main advantage of the pixelnet is that it has freedom of sampling pixel during training phase. The background of the scan is removed during training and this helps the network to converge faster. For the survival prediction task, a lot of features were studied and extracted. The features that were found to increase the error in the regression problem of survival prediction problem were removed. Various regression models were experimented and ANN was found to give the best results for the BraTS 2018 Validation data set. However, due to overfitting the same network was not able to give good results for the BraTS 2018 test data set.
\section*{Acknowledgement}
This work is supported by the Singapore Academic Research Fund under Grant {R-397-000-227-112}, NUSRI China Jiangsu Provincial Grant BK20150386 and  BE2016077 and NMRC Bedside \& Bench under grant R-397-000-245-511 awarded to Dr. Hongliang Ren.

%
%
\bibliography{mybib}{}
\bibliographystyle{splncs03}

\clearpage
\end{document}